\documentclass[preprint,aps,prb,nofootinbib,superscriptaddress]{revtex4-2}
\usepackage[utf8]{inputenc}

\usepackage{amsmath}    
\usepackage{graphicx}   
\usepackage{verbatim}   
\usepackage{color}      
\usepackage{epstopdf}   
\usepackage{upgreek} 	
\usepackage[hidelinks]{hyperref}   
\usepackage{textcomp} 	
\usepackage[separate-uncertainty=true, multi-part-units=single]{siunitx}
\usepackage{dcolumn}
\usepackage[capitalise]{cleveref}

\newcommand{\Ne}{N\'{e}el}
\newcommand{\Ga}{Ga$^{+}$}
\newcommand{\He}{He$^{+}$}
\newcommand{\ie}{\textit{i.e.}}
\newcommand{\eg}{\textit{e.g.}}

\begin{document}

\title{Local control of magnetic interface effects in chiral Ir$|$Co$|$Pt multilayers using {\Ga} ion irradiation}
\author{Mark C.H. de Jong}
\email{m.c.h.d.jong@tue.nl}
\affiliation{Department of Applied Physics, Eindhoven University of Technology, P.O. Box 513, 5600 MB Eindhoven, the Netherlands}
\author{Mari\"{e}lle J. Meijer}
\affiliation{Department of Applied Physics, Eindhoven University of Technology, P.O. Box 513, 5600 MB Eindhoven, the Netherlands}
\author{Juriaan Lucassen}
\affiliation{Department of Applied Physics, Eindhoven University of Technology, P.O. Box 513, 5600 MB Eindhoven, the Netherlands}
\author{Jos van Liempt}
\affiliation{Department of Applied Physics, Eindhoven University of Technology, P.O. Box 513, 5600 MB Eindhoven, the Netherlands}
\author{Henk J.M. Swagten}
\affiliation{Department of Applied Physics, Eindhoven University of Technology, P.O. Box 513, 5600 MB Eindhoven, the Netherlands}
\author{Bert Koopmans}
\affiliation{Department of Applied Physics, Eindhoven University of Technology, P.O. Box 513, 5600 MB Eindhoven, the Netherlands}
\author{Reinoud Lavrijsen}
\affiliation{Department of Applied Physics, Eindhoven University of Technology, P.O. Box 513, 5600 MB Eindhoven, the Netherlands}

\date{\today}

\begin{abstract}
Skyrmions are topologically protected chiral spin textures that have shown promise as data carriers in future spintronic applications. They can be stabilized by the interfacial Dzyaloshinskii-Moriya interaction (iDMI) in material systems with inversion asymmetry and spin-orbit coupling, such as Ir $|$ Co $|$ Pt multilayers. The ability to locally tune such interface interactions, and hence the skyrmion energy, could greatly enhance the nucleation and control of skyrmions in racetrack type devices. In this work, we investigate local tuning of the iDMI and perpendicular magnetic anisotropy (PMA) using focussed {\Ga} ion beam irradiation, in an Ir $|$ Co $|$ Pt multilayer system. We show that the magnitude of the interface contribution to both effects can be significantly reduced by the irradiation with {\Ga} ions. This leads to a reduction by a factor two of the domain wall energy density, while still preserving the {\Ne} character of the domain walls. Hence, we postulate that {\Ga} ion irradiation is an effective way to locally reduce the energy barrier for skyrmion nucleation, providing a novel pathway for targeted skyrmion nucleation in racetrack type devices.
\end{abstract}

\maketitle

\section{Introduction}\label{sec:Introduction}
Chiral magnetic textures such as skyrmions have shown great promise as data carriers in future spintronic memory devices \cite{Nagaosa2013, Wiesendanger2016, Fert2017, Everschor-Sitte2018, Bogdanov2020}. Their spin texture consists of a circular domain, surrounded by a domain wall with a uniform chirality. They can be very small, with diameters down to a few nanometers \cite{Buettner2018, Everschor-Sitte2018}, can be moved efficiently using electrical current pulses \cite{Woo2016} and are very stable \cite{Buettner2018}. This stability is a direct result of the uniform chirality of the domain wall \cite{Rohart2016}, which ensures that the topology of the skyrmion magnetization texture is different than the ferromagnetic background, which contributes to the energy barrier that prevents their annihilation \cite{Je2020}. These properties make chiral textures and specifically skyrmions interesting objects for future logic and data storage devices.

Chiral textures are stabilized by an asymmetric exchange interaction called the Dzyaloshinskii-Moriya interaction (DMI), which originates from the spin-orbit coupling in combination with inversion symmetry breaking \cite{Dzyaloshinskii1958, Moriya1960}. This interaction prefers a perpendicular alignment of neighbouring spins, with a well-defined chirality and hence stabilizes homochiral spin structures such as skyrmions. These have been observed in many physical systems with a DMI \cite{Everschor-Sitte2018, Kanazawa2017, Mathur2019}, but this article will focus on magnetic multilayers, in which room temperature stable {\Ne} skyrmions were first observed \cite{Moreau2016, Woo2016, Boulle2016}. In these systems the DMI originates from the interfaces between magnetic and heavy-metal layers. However, such an interface DMI (iDMI) is often not strong enough to stabilize skyrmions in single magnetic layers at room temperature. Therefore, many magnetic layers are stacked on top of each other to increase the magnetic volume, increasing the effect of the magnetic dipole field and thereby the thermal stability \cite{Moreau2016}. Many different combinations of materials have been shown to support room temperature skyrmions, {\eg} Ref. \cite{Moreau2016, Woo2016, Boulle2016, Soumyanarayanan2017, Buettner2018, Everschor-Sitte2018, Buettner2017, He2017, Legrand2017, Lemesh2018, Zhang2018, Finizio2019, Brock2020, Buettner2021}, suggesting that skyrmions can be stabilized for a wide range of magnetic parameters. These parameters can then be readily tuned by varying the layer thickness or through material choices, allowing for the optimization of the skyrmion energy cost \cite{Soumyanarayanan2017}.

However, these attributes can usually only be changed for the entire layer or device. The ability to locally control these magnetic parameters could greatly enhance the functionality of skyrmion based devices \cite{Purnama2015, Lai2017, Juge2021}, or devices that use other types of chiral textures. Enabling the creation of regions with high and low energy that could then be used to pin, guide \cite{Purnama2015, Lai2017, Juge2021} or nucleate \cite{Everschor-Sitte2017, Buettner2017, Fallon2021} such textures at desired locations in the device. It has already been well established that {\Ga} and other types of ion irradiation can be used to locally tune the magnetic parameters of single magnetic layers in a areas as small as \SI{40}{nm} \cite{Franken2011, Balk2017, Zhao2019, Diez2019}, but its effect on magnetic multilayers is not yet understood. Very recently, a study investigating the effect of low energy, broad beam {\He} ion irradiation on [Pt $|$ Co $|$ Ta]$_{\times 10}$ multilayers reported that the magnetic parameters can indeed be controlled, similar to a single magnetic layer \cite{Sud2021}. However, compared to {\He} ions, the penetration depth of {\Ga} ions is much lower \cite{Sud2021, Vieu2002} and therefore, it is not immediately obvious that {\Ga} ions can also significantly affect the effective magnetic properties of the multilayer stack.

In this Article we present the results of a systematic study on the effects of local {\Ga}ion irradiation on the magnetic parameters in an Ir $|$ Co $|$ Pt multilayer, in particular the effective anisotropy $K_{\mathrm{eff}}$ and the iDMI $D$. We will first present Hall effect measurements of the change in the effective anisotropy as a result of the {\Ga} irradiation \cite{Lavrijsen2011, Franken2014}. We find that the effective anisotropy of the stack can be readily decreased through the irradiation with {\Ga} ions, in line with work on single magnetic layers. Next, we combine these measurements with magnetic force microscopy (MFM) measurements of the stripe domain state \cite{Lemesh2017, Agrawal2019} to determine how the effective iDMI of the multilayer is affected by the {\Ga} ion irradiation and show that the iDMI also decreases in magnitude. When we only consider the interface contribution to each effect, we find that the relative decrease is the same for the anisotropy and iDMI which shows that both effects depend similarly on the interface quality and the degree of intermixing. This tuning of the magnetic parameters leads to a reduction in the domain wall energy density up to a factor 2, while still preserving the {\Ne} character of the domain walls. Hence, we postulate that ion irradiation is also an effective way to reduce and control the energy cost of different chiral structures in magnetic multilayers, such as skyrmions.

\section{Methods}\label{sec:Methods}
The complete material stack that is investigated in this article is $||$ Ta(4) $|$ Pt(2) $|$ [Ir(1) $|$ Co(0.8) $|$ Pt(1)]$_{\times 6}$ $|$ Pt(2) on top of a Si $|$ Si$\text{O}_{2}$(100) $||$ substrate. The numbers in the brackets indicate the layer thickness in \SI{}{nm}. All layers are grown using D.C. magnetron sputtering at a base pressure better than $P=$ \SI{1e-8}{\milli\bar} in an Argon atmosphere with a partial pressure of $P=$ \SI{2e-3}{\milli\bar}. The material stack is patterned into the Hall bar structures shown schematically in \cref{fig:Figure_1}(a) using standard electron beam lithography and lift-off. Following the patterning several regions of the devices are irradiated with {\Ga} ions using a FEI Nova Nanolab 600 Dualbeam operated at a beam current of \SI{1.5}{\pico\ampere} and an acceleration voltage of \SI{30}{\kilo\electronvolt}. Different regions on the devices are irradiated with different doses by varying the dwell time of the {\Ga} beam, indicated by the shaded blue regions in \cref{fig:Figure_1}(a).

To measure the local change in the anisotropy due to the ion irradiation we will determine the anisotropy from electrical Hall measurements \cite{Lavrijsen2011}. To this end, the Hall cross corresponding to this dose is electrically connected as shown in \cref{fig:Figure_1}(a). The first harmonic Hall resistance is measured by sending an AC current with a RMS current density of $j =$ \SI{1e6}{\ampere\per\meter\squared} and frequency $f = 901$ \SI{}{Hz} through the current line, the resulting Hall voltage is then measured using a lock-in amplifier.

The strength of the iDMI will be determined from measurements of the stripe domain state and a recently developed model of the equilibrium stripe domain width \cite{Lemesh2017, Agrawal2019}. These measurements are performed on the irradiated \SI{20}{\micro\meter} wide squares in between the Hall bars (labelled Islands in \cref{fig:Figure_1}(a)) with MFM, on a Br{\"u}ker Dimension Edge with custom coated low-moment tips. The irradiation of these islands was done at the same time as the Hall cross with the corresponding dose. To bring the magnetization of the multilayers into the stripe domain state, the devices shown in \cref{fig:Figure_1}(a) are demagnetized in an oscillating  magnetic field, with the field oriented approximately $85^{\circ}$ away from the sample normal \cite{Legrand2018}. The field strength starts at \SI{5}{T} and is gradually reduced by $0.5 \%$ after each oscillation, until a threshold value of 10 mT is reached. After this procedure, the magnetization inside the islands shown in \cref{fig:Figure_1}(a) exhibits a stripe domain state for all the {\Ga} doses studied. 

The saturation magnetization $M_{\text{s}}$ and the effective anisotropy $K_{\text{eff}}$ of an unpatterned sample is measured using a SQUID-VSM and the area method \cite{Johnson1996}. We find $M_{\text{s}} = 1.01 \pm 0.04$ \SI{}{MA.m^{-2}} and $K_{\text{eff}} = $ \SI{0.47 \pm 0.05}{MJ.m^{-3}}, respectively. These values are comparable to our previous work \cite{Lucassen2020}. (See Supplemental Information I).

\section{Results and discussion}\label{sec:Results}
\subsection{Anisotropy as a function of ion dose}
\begin{figure}
\includegraphics{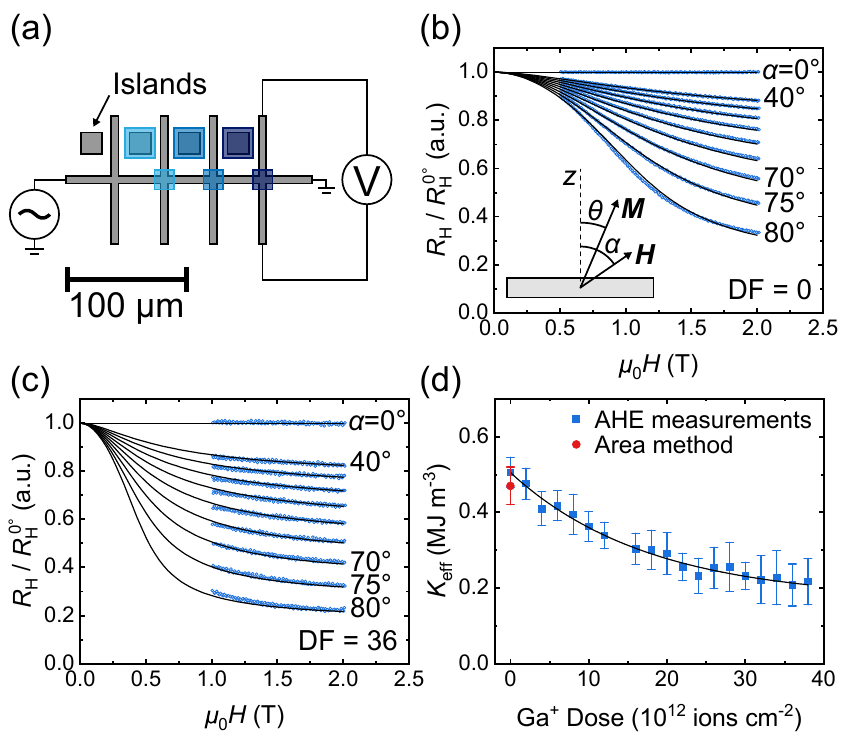}
\caption{\label{fig:Figure_1}(a) Schematic overview of the Hall bar devices. Each Hall cross has been irradiated with a different {\Ga} dose, indicated in blue. The islands in between the Hall bars are also irradiated, with the same dose as the Hall cross to their bottom right. (b) The Hall resistance measured during field sweeps from $\mu_{\mathrm{0}} H =$ \SI{2}{\tesla} to $\mu_{\mathrm{0}} H =$ \SI{0}{\tesla} (blue circles), for different angles $\alpha$, fitted using the Stoner-Wohlfarth model measured on a non-irradiated sample (black lines). Datapoints below $\mu_{\mathrm{0}} H =$ \SI{0.5}{\tesla} are not plotted to improve the clarity, but the agreement with the fit is equally good. The inset shows the definition of the angles $\alpha$ and $\theta$. DF indicates the dose-factor, \textit{i.e.} the dose is $d = \text{DF} \times 10^{12}$ ions \SI{}{cm^{-2}}. (c) The same measurement as in (b) performed on a Hall cross that has been irradiated with a dose of $d = 36 \times 10^{12}$ ions \SI{}{cm^{-2}}. The measurement is performed for $\mu_{\mathrm{0}} H =$ \SI{2}{\tesla} to $\mu_{\mathrm{0}} H =$ \SI{1}{\tesla}. (d) Plot of $K_{\mathrm{eff}}$, determined using the AHE measurements shown in (b) and (c), plotted as a function of {\Ga} dose.}
\end{figure}

We will first present the electrical Hall measurements that were used to determine the dependence of the effective anisotropy $K_{\text{eff}}$ on the {\Ga} ion dose. The measured Hall voltage is proportional to $M_{\text{z}}$, the average out-of-plane component of the magnetization $\textbf{M}$ in the region where the current line and Hall arms cross, through the anomalous Hall effect \cite{Nagaosa2010a}. During the measurement a magnetic field is applied at an angle $\alpha$ to the $z$-direction [inset in \cref{fig:Figure_1}(b)]. The effect of this field on the magnetization is to pull it away from its equilibrium out-of-plane orientation towards the in-plane direction, by an angle $\theta$. The rotation of $\textbf{M}$ is resisted by the effective anisotropy such that a stronger $K_{\text{eff}}$ results in a smaller $\theta$. This behaviour is described by the Stoner-Wohlfarth model \cite{Lavrijsen2011}.

In \cref{fig:Figure_1}(b), we have plotted the normalized first harmonic Hall resistance $R_{\text{H}}$ measured using a non-irradiated Hall cross. Shown are several field sweeps starting from $\mu_\mathrm{0}H =$ \SI{2.0}{\tesla} to $\mu_\mathrm{0}H =$ \SI{0}{\tesla}, for different angles $\alpha$ between the field and the sample normal. All measurements are normalized to the Hall resistance measured for $\alpha=0$. As expected, a larger field and angle result in a smaller $R_{\text{H}}$, since the magnetization is pulled further in-plane. The black lines in \cref{fig:Figure_1}(b) are fits to the data using the Stoner-Wohlfarth model, where $K_{\text{eff}}$ is the only fitting parameter. All field sweeps are fitted at the same time, resulting in one value for $K_{\text{eff}}$ that describes the measurements for all $\alpha$. We find a good agreement between the data and the fit. For the non-irradiated Hall cross the fit gives $K_{\text{eff}} = 0.51 \pm 0.04$ \SI{}{MJ.m^{-3}}, which is in good agreement with the value found using the area method ($K_{\text{eff}} =$ \SI{0.47 \pm 0.05}{MJ.m^{-3}}). The uncertainty in the value determined using the Hall measurements comes from the uncertainty in the value of $M_{\text{s}}$, determined independently from the SQUID-VSM measurement.

Next, we present a similar measurement on a Hall cross irradiated with a {\Ga} dose of $d = 36 \times 10^{12}$ ions \SI{}{cm^{-2}} in \cref{fig:Figure_1}(c). At this dose the sample is no longer in a single domain state at zero field and hence the Stoner-Wohlfarth model no longer describes the behaviour of the magnetization in the low field region. Therefore, we only fit the data for the part of the field sweep from $\mu_\mathrm{0}H =$ \SI{2.0}{\tesla} to $\mu_\mathrm{0}H =$ \SI{1.0}{\tesla}, where the field is strong enough to ensure a uniform magnetization, as assumed by the Stoner-Wohlfarth model. In this field range we again find good agreement between the data and the fit and determine a value for the effective anisotropy of $K_{\text{eff}} = 0.21 \pm 0.06$ \SI{}{MJ.m^{-3}}. Compared to the non-irradiated sample the anisotropy is reduced by more than a factor two. Here we have assumed that the saturation magnetization is not affected by the ion irradiation, as expected for Pt $|$ Co $|$ Pt based thin films \cite{Devolder2000}. Nevertheless, we show in Supplemental Material II that a small change in the magnetization as a function of dose will not qualitatively affect the results presented.

This procedure to measure $K_{\text{eff}}$ is performed for {\Ga} doses up to $d = 38 \times 10^{12}$ ions \SI{}{cm^{-2}} and in \cref{fig:Figure_1}(d) we plot the measured effective anisotropy as a function of {\Ga} dose. For all measurements, the AHE data was only fitted in the field range between $\mu_\mathrm{0}H =$ \SI{2.0}{\tesla} to $\mu_\mathrm{0}H =$ \SI{1.0}{\tesla}, to ensure a single domain response as required by the Stoner-Wohlfarth model. At low {\Ga} doses, between $0$ and $d = 16 \times 10^{12}$ ions \SI{}{cm^{-2}}, we observe a strong decrease in the measured anisotropy which gradually slows down for higher doses. The black line is a fit to the data and shows that the effective anisotropy decreases exponentially to a constant non-zero value as a function of the ion dose. This behaviour is consistent with previous work on Pt $|$ Co $|$ Pt single magnetic layers \cite{Vieu2002, Franken2014}, for low ion doses. The change in the anisotropy in such systems is, in part, attributed to an increase in the amount of intermixing of the heavy metal and Co atoms at the interfaces \cite{Devolder2000, Vieu2002}. This effectively makes the transition between the heavy metal layer and the Co layer smoother, which results in a more symmetric environment for the Co atoms at the interface. We expect that this will also affect the strength of the iDMI and, in particular, cause a decrease in the iDMI strength as a function of {\Ga} dose due the reduction of the inversion symmetry breaking at the interface. \textit{Ab-initio} simulations indeed predict a (small) decrease in the iDMI strength upon intermixing of a Pt $|$ Co interface \cite{Yang2015a, Zimmermann2018} and in the remainder of this Article we will investigate the dependence of the iDMI on {\Ga} ion irradiation experimentally.  

\subsection{Calculating the iDMI strength}\label{sec:MeasuringDMI}
Now that the dependence of $K_{\text{eff}}$ on the {\Ga} dose is known we will focus on determining the change in the strength of the iDMI. To do this we will use an analytical model of the magnetic stripe domain state, developed by Lemesh et al. \cite{Lemesh2017}.
This model takes as input the geometry of the multilayer stack and the magnetic parameters ($A$, $M_{\text{s}}$, $K_{\text{eff}}$, and $D$) and gives a value for the equilibrium domain width $W_{\text{eq}}$. Hence, if the equilibrium domain width is known, as well as the magnetic parameters other than the iDMI, then this model can be used to calculate the value of the iDMI strength $D$ \cite{Legrand2018, Schlotter2018, Agrawal2019}.

\begin{figure*}
\includegraphics{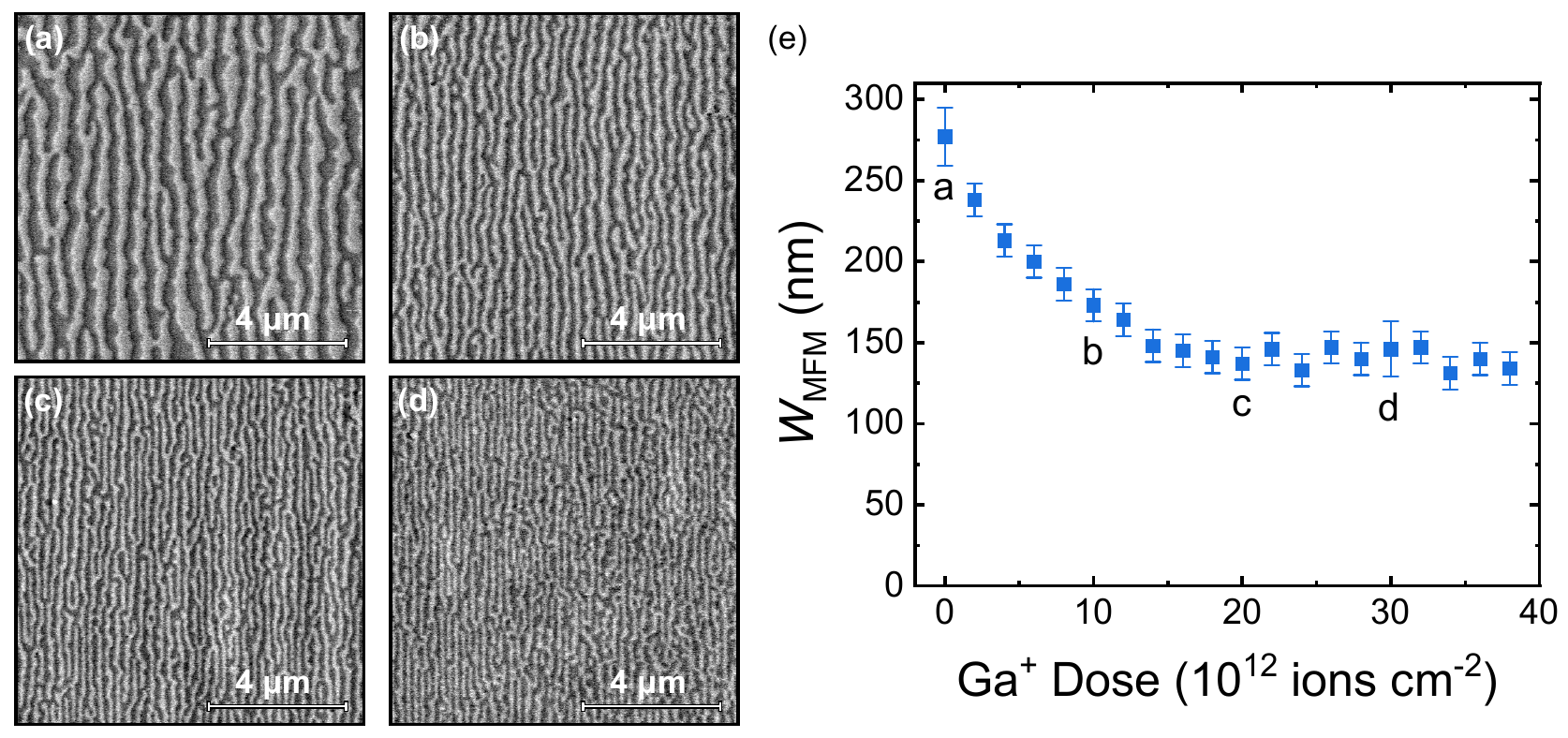}
\caption{\label{fig:Figure_2}(a - d) MFM measurements of the stripe domain state in the magnetic multilayer before and after irradiation with {\Ga} ions. The {\Ga} dose for each image is: (a) $d = 0 \times 10^{12}$ ions \SI{}{cm^{-2}}, (b) $d = 10 \times 10^{12}$ ions \SI{}{cm^{-2}}, (c) $d = 20 \times 10^{12}$ ions \SI{}{cm^{-2}}, and (d) $d = 30 \times 10^{12}$ ions \SI{}{cm^{-2}}. (e) Plot of the average domain width, determined from MFM scans, as a function of the {\Ga} ion dose.}
\end{figure*}

To this end we imaged the stripe domain state in the irradiated Ir $|$ Co $|$ Pt multilayers using MFM, as described in the Methods section. In \cref{fig:Figure_2}(a-d) we show the MFM measurements of the stripe domain state for four different {\Ga} doses, $d = 0$ ions \SI{}{cm^{-2}}, $d = 10 \times 10^{12}$ ions \SI{}{cm^{-2}}, $d = 20 \times 10^{12}$ ions \SI{}{cm^{-2}}, and $d = 30 \times 10^{12}$ ions \SI{}{cm^{-2}} for (a) - (d), respectively. The scale bar for all four images is identical and hence it is immediately clear that {\Ga} irradiation has a strong effect on the domain size. For increasing dose we observe a strong decrease of the domain width. To quantify this change we measure the domain width in each MFM scan using the 2D Fourier transform (described in detail in Ref. \cite{Lucassen2017}). We confirmed that the equilibrium domain size in the non-irradiated islands is the same as the measured domain size in an unpatterned sample (not shown).

The measured domain size $W_{\text{MFM}}$ is plotted against the {\Ga} dose in \cref{fig:Figure_2}(e). Qualitatively, the observed trend in the domain size is similar to the observed trend for the effective anisotropy. Until a {\Ga} dose of approximately $d = 12 \times 10^{12}$ ions \SI{}{cm^{-2}} the domain width decreases rapidly as a function of dose. This decrease slows down for higher doses and eventually saturates around $W_{\text{MFM}} = 140 \pm 2$ \SI{}{nm}. The decrease in the domain size can be understood by considering the domain wall energy density \cite{Thiaville2012},
\begin{equation}\label{eq:DW_energy}
\sigma_{\text{DW}} = 4 \sqrt{A K_{\text{eff}}} - \pi |D|.
\end{equation}

As the effective anisotropy decreases the energy cost of a domain wall also decreases, resulting in an increase in the number of domain walls and narrower domains. Comparing \cref{fig:Figure_1}(d) and \cref{fig:Figure_2}(e) we see that the dependence of $K_{\text{eff}}$ and $W_{\text{MFM}}$ on the {\Ga} dose differ for high doses ({\ie} $d > 20 \times 10^{12}$ ions \SI{}{cm^{-2}}). The domain width remains constant while the anisotropy continues to decrease, suggesting that the magnitude of the iDMI should also decreases in this regime to keep $\sigma_{\text{DW}}$ constant.

To quantify this change we use the aforementioned model to calculate the value of the iDMI using the measured values for $K_{\text{eff}}$ and $W_{\text{MFM}}$, for the saturation magnetization we use $M_{\text{s}} = 1.01 \pm 0.04$ \SI{}{A.m^{-2}}, as measured for the non-irradiated sample and for the exchange stiffness we use a value of $A = 10$ \SI{}{pJ.m^{-1}}, in accordance with other work on similar multilayers \cite{Moreau2016, Legrand2018}. In Supplemental Material III we show that the value of $A$ does not impact our findings qualitatively, only quantitatively. Uncertainties in the value of $D$ are calculated using the same procedure as in Ref. \cite{Agrawal2019}.

\begin{figure*}
\includegraphics{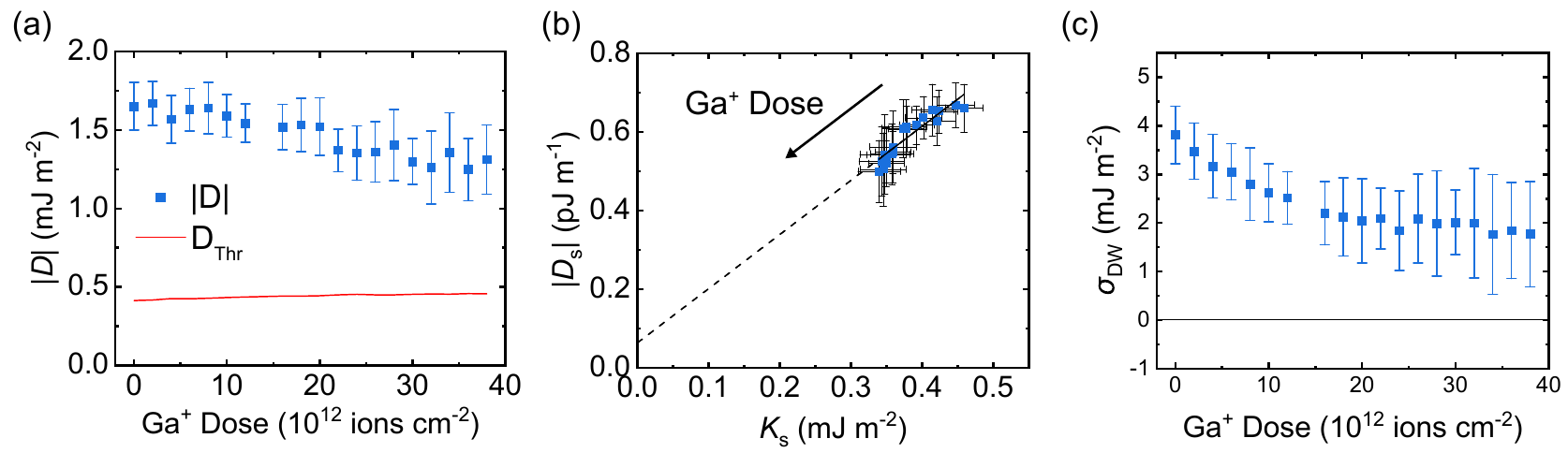}
\caption{\label{fig:Figure_3}(a) Plot of the calculated magnitude of the iDMI as a function of the {\Ga} dose (blue). The red line is the threshold iDMI value $D_{\text{thr}}$ above which the domain walls are of the {\Ne} type. (b) The interface DMI $D_{\text{s}}$ plotted as a function of the interface anisotopy $K_{\text{s}}$. In the measured dose range, the relative decrease in $D_{\text{s}}$ and $K_{\text{s}}$ is approximately the same. (c) The domain wall energy density $\sigma_{\text{DW}}$ plotted as a function of the {\Ga} dose.}
\end{figure*}

The magnitude of the iDMI is plotted as a function of the {\Ga} dose in \cref{fig:Figure_3}(a). Without {\Ga} irradiation we measure an iDMI of $|D| = 1.7 \pm 0.2$ \SI{}{mJ.m^{-2}}, which is consistent with measurements performed by other groups on similar material stacks \cite{Legrand2018, Schlotter2018}. We observe that the strength of the iDMI gradually decreases over the entire dose range studied here, down to $|D| = 1.3 \pm 0.2$ \SI{}{mJ.m^{-2}} for the largest {\Ga} dose ($d = 38 \times 10^{12}$ ions \SI{}{cm^{-2}}). This is consistent with the hypothesis that the increase in the degree of intermixing is responsible for the observed changes in $K_{\text{eff}}$ and $D$. The red line in \cref{fig:Figure_3}(a) corresponds to the minimum iDMI strength $D_{\text{Thr}}$ required to ensure the formation of N{\'e}el walls over Bloch walls \cite{Lemesh2017}. We find that the measured iDMI values are at least a factor 2 greater than this threshold, for all doses, indicating that the irradiation will not affect the chirality of the domain walls in the multilayer\footnote{This threshold value does not include the effect of hybrid chiralities due to the dipolar interactions \cite{Dovzhenko2018, Legrand2018, Lemesh2018a}, which might affect the chirality of the domain walls. In Supplemental Material V we show using MuMax$^{3}$ \cite{Vansteenkiste2014} that it is likely that the chirality in the multilayer is still uniform after irradiation, even at the highest dose used, $d = 38 \times 10^{12}$ ions \SI{}{cm^{-2}}.}.

Next, we compare the observed change in the anisotropy and iDMI. To this end we calculate the interface contribution to each effect, $K_{\text{s}}$ and $D_{\text{s}}$ for the anisotropy and DMI, respectively. The conversion is done using the following relations,
\begin{align}
K_{\text{eff}} = \frac{2 K_{\text{s}}}{t_{\text{Co}}} - \frac{1}{2} \mu_{\text{0}} M_{\text{s}}^{2},\\
\nonumber \\
D = \frac{2 D_{\text{s}}}{t_{\text{Co}}}.
\end{align}
The factor two in front of the interface contribution in each equation accounts for the fact that there are two heavy metal interfaces with the Co layer. This means that $K_{\text{s}}$ and $D_{\text{s}}$ are the average interface contributions of the Ir and Pt interfaces, as we have no method to distinguish between the two.

In \cref{fig:Figure_3}(b) we plot the interface contribution to the iDMI against the interface contribution to the anisotropy. Increasing the {\Ga} dose results in a decrease of both $D_{\text{s}}$ and $K_{\text{s}}$ as indicated by the arrow. The data is fitted with a straight line (solid black line), which fits the data well in the studied dose range. Extrapolating this fit towards lower values (dashed black line), we find that it passes through the origin within the experimental uncertainty $D_{\text{s}}(K_{\text{s}} = 0) = 0.1 \pm 0.5$ \SI{}{pJ.m^{-1}}. This shows that the relative decrease in $D_{\text{s}}$ and $K_{\text{s}}$ is the same upon irradiation with {\Ga} ions, suggesting that the dependence on the degree of intermixing at the interface is the same for both effects (in the studied dose range). This result is in line with earlier studies, where the interfaces responsible for the iDMI and anisotropy are modified using annealing \cite{Kim2019}, {\He} ion irradiation on single magnetic layers \cite{Zhao2019} or through the crystal phase \cite{Kim2018}. In all cases, a correlation between the interface contributions to the anisotropy and iDMI was reported.

The effect of the these changes in the anisotropy and iDMI on the domain wall energy is shown in \cref{fig:Figure_3}(c). Here we plot the domain wall energy density, calculated using \cref{eq:DW_energy}, as a function of the {\Ga} dose. In the studied dose range, we can reduce the domain wall energy density by a factor 2. Taken together with the fact that the iDMI always remains larger than the threshold value for {\Ne} type domain walls (\cref{fig:Figure_3}(a)), we conclude that {\Ga} ion irradiation is an effective way to decrease the energy cost of chiral domain walls. Decreasing the domain wall energy also decreases the energy of chiral textures such as skyrmions, which has been shown to result in more efficient field driven nucleation \cite{Soumyanarayanan2017}. Hence, we conjecture that {\Ga} ion irradiation can also be used to locally tune the energy and properties of chiral magnetic textures such as skyrmions in multilayer systems.

\section{Discussion and Conclusion}
We have shown that magnetic effects with an interface origin can be locally modified using a {\Ga} ion beam. In our analysis we have made several assumptions that are relevant for the interpretation of the obtained results. Here we explicitly list these assumption and discuss their consequences. (i) We assumed that the saturation magnetization is not affected by ion irradiation. This assumption is based on early work on the effect of {\Ga} ion irradiation of Pt $|$ Co $|$ Pt single magnetic layers. For comparable ion doses as those used in this Article, either no or a small ($< 5\%$) change in $M_{\text{s}}$ is reported \cite{Devolder2000, Vieu2002}. Both a decrease and increase could in theory occur, due to the intermixing of Co and Pt \cite{Vieu2002}. In Supplemental Material II we show that a small decrease or increase in $M_{\text{s}}$ does not influence the obtained results significantly. (ii) We assumed that the value of the exchange stiffness is equal to $A = 10$ \SI{}{pJ.m} and is not affected by the ion irradiation. The choice for the value of $A$ is based on other work on similar magnetic multilayers \cite{Moreau2016, Legrand2018, Lucassen2020}. In Supplemental Material III we show that using a different value for $A$ does not qualitatively affect the results. A change in the exchange stiffness due to ion irradiation has not been reported to the best of our knowledge and is not considered here. (iii) Finally, we did not directly consider the depth dependence of the ion irradiation. Contrary to lighter {\He} ions, {\Ga} ions have a significantly lower penetration depth resulting in a depth dependent damage profile \cite{Sud2021, Vieu2002}. The number of layers of the magnetic multilayer was chosen to ensure that all the layers are affected by the ion irradiation to some extent, as can be seen from the TRIM \cite{Ziegler2010} simulations in Supplemental Material IV, while maximizing the number of layers to ensure {\eg} skyrmion stability. Although this depth dependence will mean that the magnetic parameters become depth dependent, the measurements reported in this Article measure the effective anisotropy and effective iDMI, which correspond to the layer averaged values of these parameters. In the case of the anisotropy measurement this is straightforward to see, since the Hall signal is proportional to the average $M_{\text{z}}$ inside the Hall cross. In Supplemental Material V we show that this is also the case for the measurements of the iDMI.

To conclude, in this work we have investigated local tuning of the interface DMI and perpendicular magnetic anisotropy using {\Ga} ion irradiation, in an Ir $|$ Co $|$ Pt multilayer system. We showed that irradiation with {\Ga} ions has a significant effect on the interface contributions to both effects. The net effect of this is to reduce the energy cost of domain walls by up to a factor 2, while still preserving their chiral {\Ne} character. Hence, we postulate that {\Ga} ion irradiation is an effective way to locally---with a resolution of $\sim$\SI{40}{nm}---reduce the energy barrier for the nucleation of skyrmions and other chiral textures. Providing a novel pathway towards the control of chiral textures in future spintronic devices.

\begin{acknowledgments}
This work is part of the Gravitation programme 'Research Centre for Integrated Nanophotonics', which is financed by the Netherlands Organisation for Scientific Research (NWO). M. J. M. and J. L. acknowledge support as part of the research programme of the Foundation for Fundamental Research on Matter (FOM), which is a part of NWO.
\end{acknowledgments}

\nocite{NoteX}

\end{document}


\title{Supplemental Material: Local control of magnetic interface effects in chiral Ir$|$Co$|$Pt multilayers using {\Ga} ion irradiation}
\author{Mark C.H. de Jong}
\email{m.c.h.d.jong@tue.nl}
\affiliation{Department of Applied Physics, Eindhoven University of Technology, P.O. Box 513, 5600 MB Eindhoven, the Netherlands}
\author{Mari\"{e}lle J. Meijer}
\affiliation{Department of Applied Physics, Eindhoven University of Technology, P.O. Box 513, 5600 MB Eindhoven, the Netherlands}
\author{Juriaan Lucassen}
\affiliation{Department of Applied Physics, Eindhoven University of Technology, P.O. Box 513, 5600 MB Eindhoven, the Netherlands}
\author{Jos van Liempt}
\affiliation{Department of Applied Physics, Eindhoven University of Technology, P.O. Box 513, 5600 MB Eindhoven, the Netherlands}
\author{Henk J.M. Swagten}
\affiliation{Department of Applied Physics, Eindhoven University of Technology, P.O. Box 513, 5600 MB Eindhoven, the Netherlands}
\author{Bert Koopmans}
\affiliation{Department of Applied Physics, Eindhoven University of Technology, P.O. Box 513, 5600 MB Eindhoven, the Netherlands}
\author{Reinoud Lavrijsen}
\affiliation{Department of Applied Physics, Eindhoven University of Technology, P.O. Box 513, 5600 MB Eindhoven, the Netherlands}

\date{\today}

\maketitle
\section{Squid measurements on unpatterned [Ir(1.0)$|$Co(0.8)$|$Pt(1.0)]$_{\times 6}$}
\begin{figure*}[h]
\includegraphics{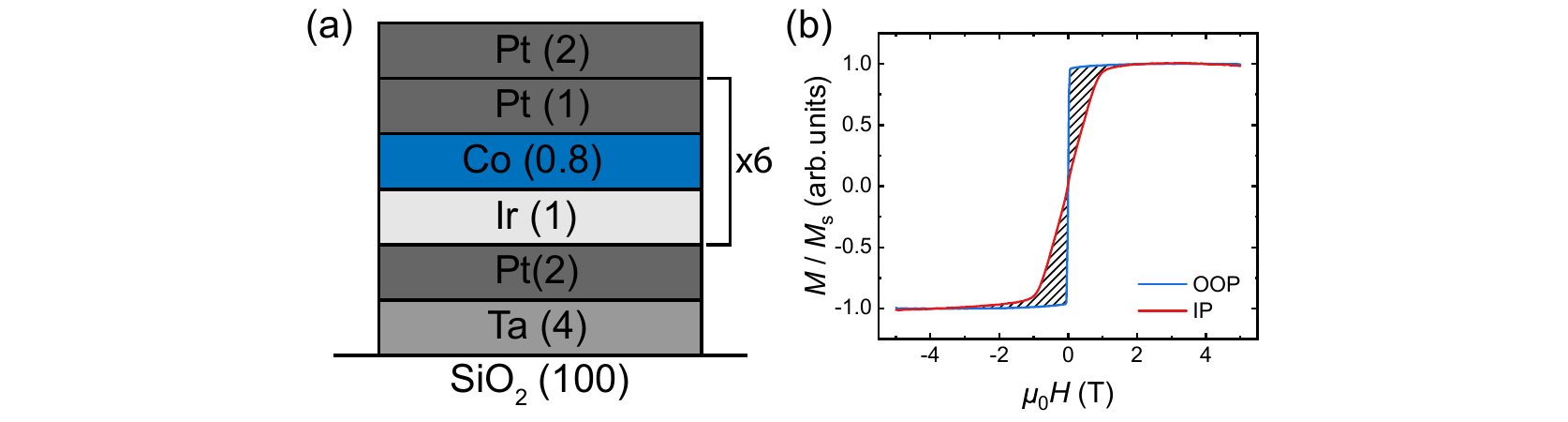}
\caption{\label{fig:Figure_1}(a) Schematic overview of the sample stack. (b) SQUID-VSM measurements of a $4 \times 4$ \SI{}{mm^{2}} sample of the non-irradiated multilayer shown in (a). $M_{\text{s}}$ was measured to be $M_{\text{s}} = 1.01 \pm 0.04$ \SI{}{MA.m^{-1}}, to convert the measured moment $|\textbf{m}|$ (in \SI{}{A.m^{2}}) to a magnetization we assumed that the total thickness of the magnetic volume is equal to the total Co thickness, $6 \times 0.8$ \SI{}{nm}. The area between the in-plane (IP) and out-of-plane (OOP) loop (shaded with black lines) was used to calculate the effective anisotropy using the area method \cite{Johnson1996}. We find $K_{\text{eff}} = 0.47 \pm 0.05$ \SI{}{MJ.m^{-3}}.}
\end{figure*}

\newpage
\section{The effect of small changes in the saturation magnetization}
\begin{figure*}[h]
\includegraphics{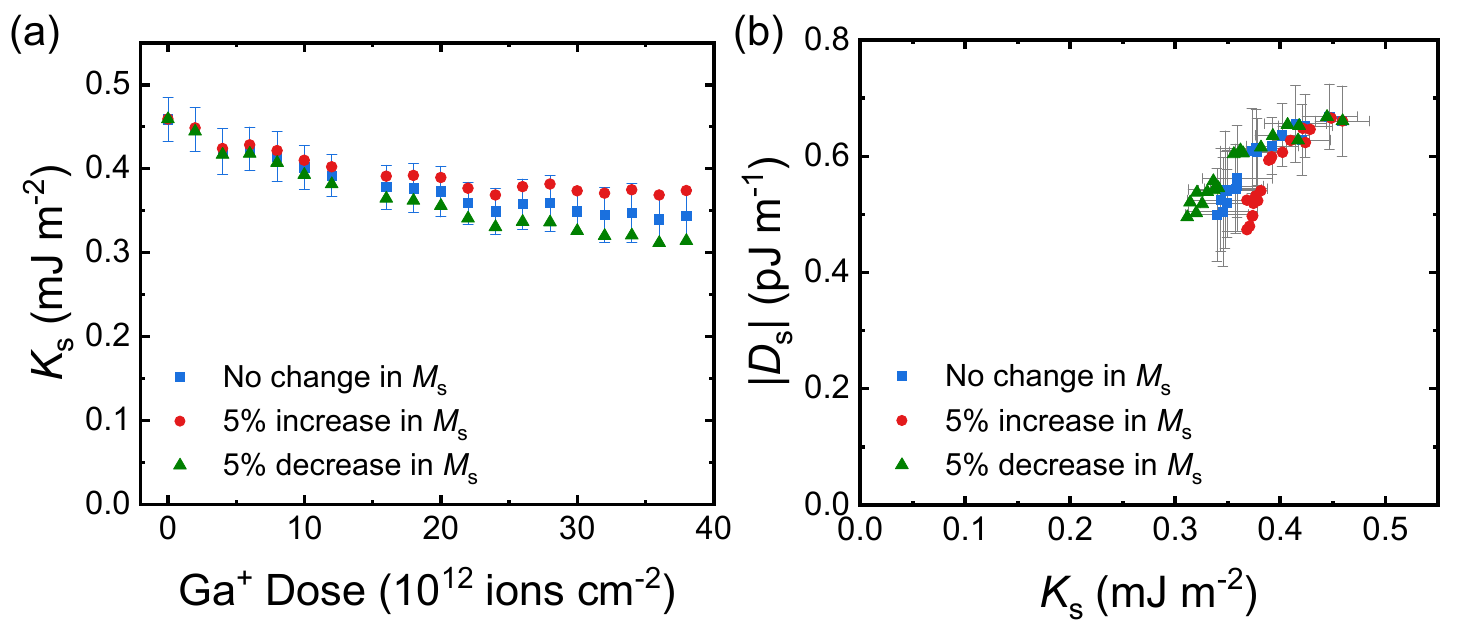}
\caption{\label{fig:Figure_3} (a) Plot of the interface anisotropy $K_{\text{s}}$ as a function of ion dose. If $M_{\text{s}}$ changes gradually as a function of dose, then the value of interface anisotropy determined using our measurement will change as well. (b) $|D_{\text{s}}|$ plotted as a function of $K_{\text{s}}$ for constant (blue) and changing $M_{\text{s}}$ (red and green). The error bars for the red and green data points are not shown in both figures, but are comparable in size to the blue dataset.}
\end{figure*}

In the main paper it was assumed that the saturation magnetization stays constant as a function of ion dose. This corresponds to measurements performed by Devolder \cite{Devolder2000}. However, Vieu et al. \cite{Vieu2002} estimate that small changes in $M_{\text{s}}$ are possible for a system consisting of Pt$|$Co$|$Pt due to the formation of a Pt$/$Co alloy at larger doses. They find that $M_{\text{s}}$ can change (increase or decrease) by approximately $5{\%}$ for the maximum dose used in the main text. In \cref{fig:Figure_3} we show how our results would be affected by such a change. We assume here that $M_{\text{s}}$ depends linearly on the dose.

The most significant effect is that the value found for the interface anisotropy will change, although the observed decrease upon irradiation is still clearly present as seen in \cref{fig:Figure_3}(a). When $M_{\text{s}}$ is increased, the Zeeman term in the Stoner-Wohlfarth model increases. Hence, to reproduce the experimentaly observed behaviour, the anisotropy term must also increase. Indeed, for an increasing $M_{\text{s}}$ we find an increasing $K_{\text{s}}$ and \textit{vice versa}, compared to a constant $M_{\text{s}}$. In \cref{fig:Figure_3}(b), we plot the interface DMI versus the interface anisotropy, for constant, increasing and decreasing $M_{\text{s}}$. Here the behaviour of $M_{\text{s}}$ can slightly change the relative scaling of the two parameters as a function of dose, resulting in a faster or slower decrease in $D_{\text{s}}$ compared to $K_{\text{s}}$. Nevertheless, this effect remains small and all values fall within the error bars shown in the main text.

\section{The value of the exchange stiffness}
\begin{figure*}[h]
\includegraphics{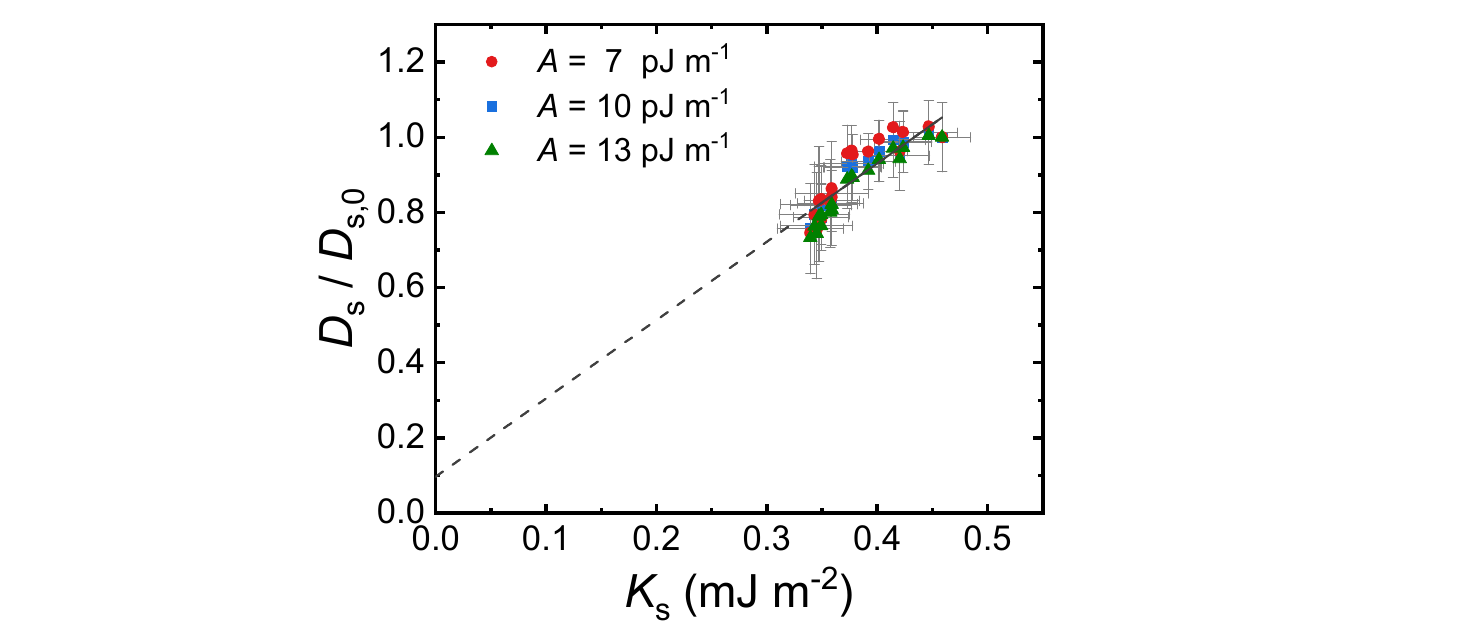}
\caption{\label{fig:Figure_2} Plot that shows how the results presented in Figure 3(b) depend on the value of the exchange stiffness $A$. This plot is equivalent to Figure 3(b) in the main text, with the only difference that we now plot the normalized interface DMI on the vertical axis. Each data set is normalized to the corresponding $D_{\text{s}}$ for zero dose. When this is done for three different values of the exchange stiffness, we find that all the data overlap. In the main paper the value $A =$ \SI{10}{pJ.m^{-1}} was used. We conclude that the value of the iDMI $D$ depends on $A$, but the effect of the ion irradiation reported in the main paper is independent of the value of $A$. The linear fit to the data and the error bars both correspond to the dataset for $A =$ \SI{10}{pJ.m^{-1}}.}
\end{figure*}

\newpage

\section{Depth dependence of the {\Ga} ion irradiation}\label{sec:Irradiation}
\begin{figure*}[h]
\includegraphics{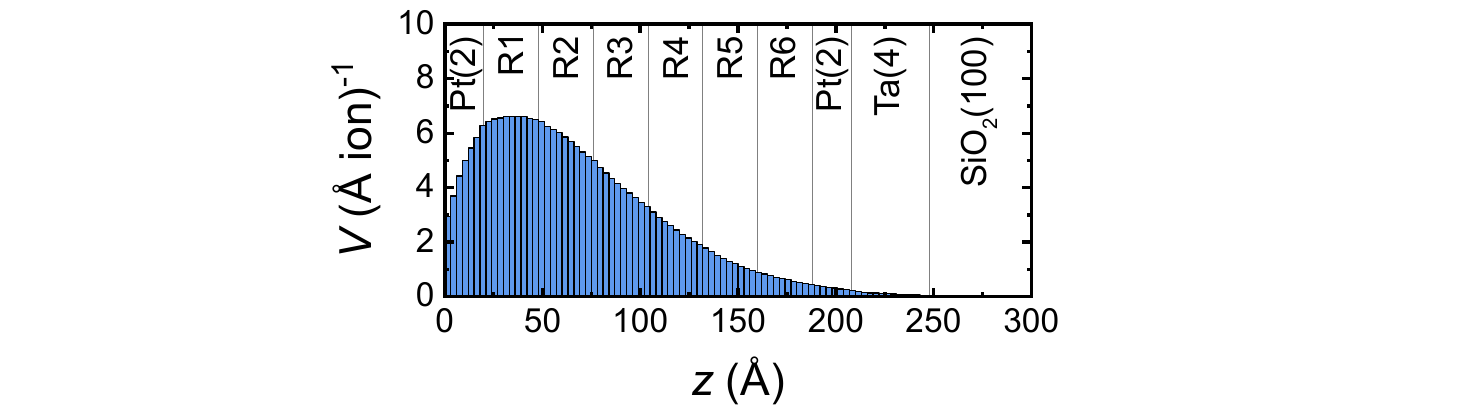}
\caption{\label{fig:Figure_4} Plot of the average number of vacancies created by each ion during the irradaition for a sample with 6 repeats, labelled R1 through R6. Each repeat consists of a 2.8 nm thick Ir $:$ Co $:$ Pt alloy with a stochiometric ratio of 10 $:$ 8 $:$ 10. Simulations are performed using the TRIM code with a total of 50,000 ions with an energy of \SI{30}{keV}.}
\end{figure*}

The penetration depth of {\Ga} ions into Pt and Co is on the order of \SI{10}{nm} \cite{Hyndman2001}. Since the thickness of our magnetic multilayer system is larger, we expect that the effect of the ion irradiation is not uniform throughout the thickness of the multilayers. To get an estimate of the damage profile as a function of thickness, we have used the Transport and Ranges of Ions in Matter (TRIM) code \cite{Ziegler2010} to simulate the effect of the impinging {\Ga} ions on the multilayers. The material stack that is implemented in the software is SiO$_{2}|$Ta(4)$|$Pt(2)$|$[R]$_{\times 6}|$Pt(2). Here R denotes the magnetic repeat Ir(1)$|$Co(0.8)$|$Pt(1). Due to limitations in the maximum layer number within the software we were forced to model the repeat as an Ir $:$ Co $:$ Pt alloy, with a stochiometric ratio of $10:8:10$. In total 50,000 collision events are simulated, with \SI{30}{keV} {\Ga} ions, corresponding to the irradiation applied in the experiment. In \cref{fig:Figure_4} we plot the average number of vacancies (dislocated atoms) created by each ion as a function of the depth into the magnetic multilayer. From this plot it is clear that there is a thickness dependence, with the top layers being influenced more strongly than the bottom layers. Based on these simulations the number of magnetic layers was limited to 6 in the main paper, to ensure that all layers are affected by the ion irradiation. Despite the thickness dependence, the results in the main paper show that the average magnetic interface properties of the stack are strongly influenced by the ion irradiation. In \cref{sec:MuMax} of the Supplemental material we show using MuMax$^{3}$ that the measurement of the iDMI is still valid, even if there is a layer dependence in both the anisotropy and iDMI.

\newpage

\section{MuMax$^{3}$ simulations of non-uniform irradiation}\label{sec:MuMax}
Using MuMax$^{3}$ \cite{Vansteenkiste2014} we performed micromagnetic simulations of the magnetic multilayer stack after ion irradiation. Here we investigate the correctness of the assumption that a measurement of the iDMI using the domain width in a stack in which the uniaxial anisotropy and iDMI vary as a function of layer number will result in the average iDMI parameter of the layers. The approach we take is similar to the approach used in Ref. \cite{Lucassen2020}, to verify their 'averaging approach'. The material system we simulate is $[\text{NM}(2)/\text{FM}(1)]\times N$, where $N =6$ is the total number of repeats of the \SI{2}{nm} thick nonmagnetic (NM) layer and \SI{1}{nm} thick ferromagnetic (FM) layer. The average magnetic parameters are chosen to be equal to those of the stack irradiated with a {\Ga} dose of $d = 12 \times 10^{12}$ ions \SI{}{cm^{2}}: $M_{\text{s}} =$ \SI{1.0}{MA.m^{-1}}, $A =$ \SI{10}{pJ.m^{-1}}, $K_{\textrm{u}} =$ \SI{0.98}{MJ.m^{-3}}, and $D_{\text{avg}} =$ \SI{1.5}{mJ.m^{-2}}. In the case of a non-uniform stack, both the uniaxial anisotropy and iDMI are layer dependent and vary linearly as a function of layer number (to approximate the depth dependence in \cref{fig:Figure_4}). The difference between two successive layers is $\Delta K_{\text{u}} =$ \SI{0.05}{MJ.m^{-3}} and $\Delta D =$ \SI{0.05}{mJ.m^{-2}}, the average value of both parameters is equal to the corresponding value in the uniform stack. These differences between successive layers were chosen so that the value of $D$ and $K_{\text{u}}$ in the bottom layer are approximately the same as in the non-irradiated stack. Representing a scenario where the sixth layer is not affected by the irradiation, {\ie} the largest possible gradient.
 
\begin{figure*}[h]
\includegraphics{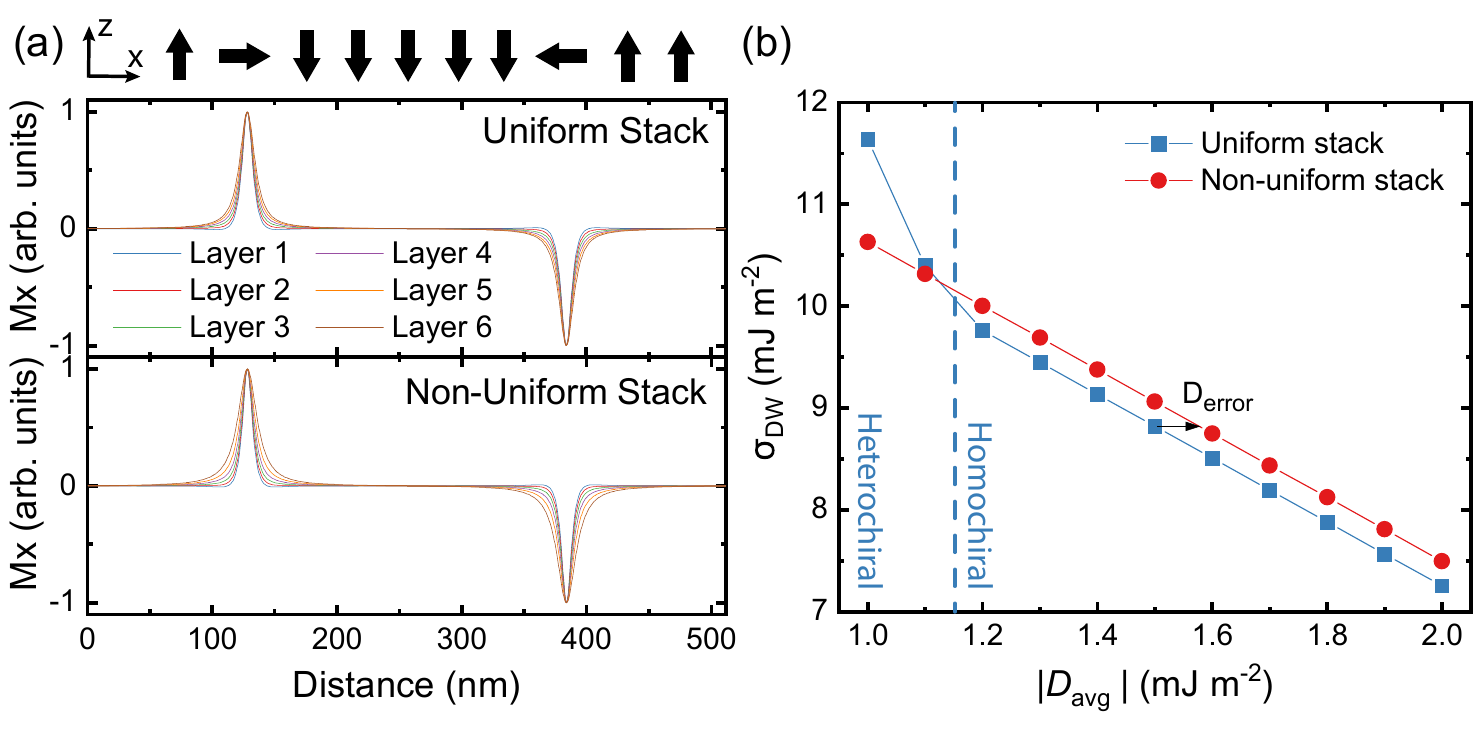}
\caption{\label{fig:Figure_5}(a) The $x$-component of the magnetization in each of the 6 repeats, plotted for both the uniform and non-uniform stack. In both cases the domain walls have a counter-clockwise chirality in all of the layers, for an average iDMI constant of $D_{\text{avg}} =$ \SI{1.5}{mJ.m^{-2}} (illustrated by the arrows at the top of the figure). (b) Simulated domain wall energy density in the uniform and non-uniform systems. Above $|D_{\text{avg}}| =$\SI{1.2}{mJ.m^{-2}}, a linear dependence on $D_{\text{avg}}$ is observed in both cases. Below this value the domain walls in the uniform stack no longer have the same chirality, resulting in a deviation from this linear dependence. In the non-uniform stack, the average domain wall energy density is slightly lower. The model of Lemesh \cite{Lemesh2017} models a uniform stack, and therefore slightly overestimates the iDMI by $D_{\text{error}} \approx$ \SI{0.08}{mJ.m^{-2}}, which falls well within the experimental uncertainty for $D$.}
\end{figure*}

We simulate a region of $512$ by $32$ \SI{}{nm^{2}} with periodic boundary conditions in the x and y direction of 32 repeats. The cell sizes are given by $(x, y, z) = (0.5, 8, 1)$ \SI{}{nm}. Two domain wall are then initialized in these systems, with the domain wall normal along the $x$-direction. The width of these initial walls is set to \SI{5}{nm} and the domain wall moment is set to $45^{\circ}$ from the domain wall normal. The systems are subsequently relaxed, resulting in chiral {\Ne} walls in all layers for both the uniform and non-uniform stack, as shown in \cref{fig:Figure_5}(a). This is important because the model of Lemesh \cite{Lemesh2017} assumes that the domain wall profile in each layer is the same. We also verified that for the largest ion dose used in this Article, $d = 38 \times 10^{12}$ ions \SI{}{cm^{-2}}, both scenarios results in CW {\Ne} walls in all layers. 

Next, we focus on extracting the domain wall energy density, to verify that the model of Lemesh can still be used in the case of a gradient in $K_{\textrm{u}}$ and $D$. We calculate the domain wall energy density by comparing the energy of the systems with domain walls to a uniformly magnetized system, the result is plotted in \cref{fig:Figure_5}(b). Above $|D_{\text{avg}}| =$\SI{1.2}{mJ.m^{-2}}, a linear dependence of the domain wall energy on $D_{\text{avg}}$ is observed in both cases. In the non-uniform stack, the domain wall energy density is slightly lower than in the uniform stack. The domain width in our measurements is determined by the domain wall energy and hence we look at the difference in $D_{\text{avg}}$ that is present when the two systems have the same domain wall energy density. Since the model of Lemesh \cite{Lemesh2017} assumes a uniform stack, this means that the iDMI is overestimated by $D_{\text{error}} \approx$ \SI{0.08}{mJ.m^{-2}} in our measurements, which falls entirely in our experimental uncertainty. Hence the use of the model of Lemesh is a valid approach to calculate the average $D$ in our system, even if a layer dependence of the magnetic parameters is present.

Below $|D_{\text{avg}}| =$\SI{1.2}{mJ.m^{-2}}, the domain walls in the uniform stack no longer all have a CW chirality. The domain wall moments in the bottom of the stack rotate to align with the dipolar field, since the DMI is no longer strong enough to prevent this. This causes an increase in the domain wall energy density plotted in \cref{fig:Figure_5}(b), because we do not include the dipolar energy. Interestingly, in the non-uniform scenario the domain walls in the simulation all kept their CW chirality. This can be understood because the DMI in the bottom layers is larger than in the top layers in this scenario and thus still strong enough to overcome the dipolar field. This suggests that a depth dependent irradiation profile ({\Ga} irradiation) might be beneficial over a uniform profile ({\He} irradiation) for multilayers, if the uniform chirality is important for the application.